\documentclass[]{spie}  

\usepackage{amsmath,amsfonts,amssymb}
\usepackage{graphicx}
\usepackage{braket}
\usepackage{siunitx}
\usepackage[colorlinks=true, allcolors=blue]{hyperref}

\title{Towards an Engineering Framework for Ultrafast Quantum Nonlinear Optics}
\author[*,a]{Ryotatsu Yanagimoto}
\author[*,a]{Edwin Ng}
\author[b,c]{Tatsuhiro Onodera}
\author[a]{Hideo Mabuchi}
\affil[a]{E.\,L.\,Ginzton Laboratory, Stanford University, Stanford, California 94305, USA}
\affil[b]{NTT Physics and Informatics Laboratories, NTT Research, Inc., 1950 University Ave., East Palo Alto, California 94303, USA}
\affil[c]{School of Applied and Engineering Physics, Cornell University, Ithaca, New York 14853, USA}

\authorinfo{Further author information: (Send correspondence to H.M.)\\R.Y.: ryotatsu@stanford.edu\\
E.N.: edwin98@stanford.edu\\
T.O.: tatsuhiro.onodera@ntt-research.com\\
H.M.: hmabuchi@stanford.edu\\
*R.Y. and E.N. equally contributed to this work.}

\begin{document} 
\maketitle

\begin{abstract}
The advent of dispersion-engineered and highly nonlinear nanophotonics is expected to open up an all-optical path towards the strong-interaction regime of quantum optics by combining high transverse field confinement with ultra-short-pulse operation. Obtaining a full understanding of photon dynamics in such broadband devices, however, poses major challenges in the modeling and simulation of multimode non-Gaussian quantum physics, highlighting the need for sophisticated reduced models that facilitate efficient numerical study while providing useful physical insight. In this manuscript, we review our recent efforts in modeling broadband optical systems at varying levels of abstraction and generality, ranging from multimode extensions of quantum input-output theory for sync-pumped oscillators to the development of numerical methods based on a field-theoretic description of nonlinear waveguides. We expect our work not only to guide ongoing theoretical and experimental efforts towards next-generation quantum devices but also to uncover essential physics of broadband quantum photonics. 
\end{abstract}


\section{Introduction}
\label{sec:intro}
One of the major obstacles to achieving quantum engineering and information processing of light on all-photonic platforms is the difficulty of realizing strong coherent nonlinearities in optical materials. The high bandwidth, low power, and room-temperature operability of laser-based coherent optical devices have made such systems ubiquitous in the fields of communications~\cite{Marin-Palomo2017}, advanced/enhanced sensing~\cite{LIGO2013, Coddington2016}, and precision metrology~\cite{Udem2002}. But in order to bring all-optical platforms into the domain of quantum-scale engineering currently occupied by more specialized hardware such as superconducting circuits~\cite{Devoret2013, Barends2013}, it is necessary, for example, for the presence of only a few photons in one optical channel to condition or trigger a coherent manipulation of photons in a neighboring optical channel~\cite{Chang2014, Birnbaum2005}. Previous generations of proof-of-concept experiments using continuous-wave light interacting strongly with atomic nonlinearities have shown unequivocally that quantum-scale engineering is possible in principle when the rate of nonlinear interactions $g$ and the typical decoherence rate $\kappa$ of the system enter the regime of \emph{strong coupling} where the figure of merit such as $g/\kappa \sim 1$~\cite{Turchette1995, Mabuchi2002, Birnbaum2005}. In the absence of strong coupling, access to non-classical physics in optical hardware is limited to schemes based on, e.g., measurement and feedback, which result in stochastic operation~\cite{Knill2002, Miyata2016, Knill2001, Bartlett2002}. Thus, an important direction for next-generation quantum-optical research is the translation of these quantum capabilities into all-photonic hardware which can access the quantum nature of light while also being amenable to technological scaling, e.g., via chip integration. To that end, one of the most promising ways of tackling this problem is to step away from the continuous-wave paradigm of traditional quantum optics and instead leverage the rich phenomenology and features inherent in \emph{multimode} and \emph{broadband} optical fields which have recently become experimentally accessible due to advances in the field of ultrafast optics~\cite{Weiner2011}.

Ultrafast light sources in the form of mode-locked pulsed lasers and optical frequency combs have become a major driving force behind recent advances in precision coherent optical techniques, such as frequency \cite{Udem2002} and distance \cite{Thiel2017} metrology, dual-comb spectroscopy \cite{Coddington2016}, mid-IR tunable light generation \cite{Marandi2016}, and efficient nonlinear optics~\cite{Boyd2008}. The phase-coherent excitation of light across many THz of bandwidth constitutes, in the time domain, a pulsed output featuring extremely high peak power, and the concentration of electric field density into ultrashort, often femtosecond-scale, pulses can greatly enhance the nonlinear response of a material which would otherwise only weakly interact with a continuous-wave laser beam. Using nonlinear optical processes such as three-wave mixing in $\chi^{(2)}$ crystals like periodically-poled lithium niobate, it is possible to manipulate information stored within the state of the multimode pulses~\cite{Roslund2014}, and the significant enhancement in nonlinearity afforded by the high peak power opens the possibility of pushing this control into the few-photon regime where quantum effects become important. Notably, the last few years have witnessed significant progress in the fabrication of ultra-low loss and highly nonlinear nanophotonic platforms~\cite{Wang2018, Stanton2020}, where $g/\kappa\sim\num{0.01}$ has already been achieved in continuous-wave systems~\cite{Lu2020}. The fact that some of these very same systems can additionally support ultra-short pulse operation via advanced dispersion engineering~\cite{Jankowski2020}, then, gives significant weight to the potential of coherently generating and manipulating non-Gaussian states of light in these platforms within the near future~\cite{Yanagimoto2020}.

To fully leverage the technological capabilities of current and future quantum optical devices, however, we need to overcome major theoretical and modeling difficulties. In the strongly interacting regime of broadband optics, both the multimode and the ``discrete'' nature of photons have to be properly captured, rendering na\"ive quantum models prohibitively expensive numerically. These challenges underscore the importance of developing sophisticated reduced models that are capable of enabling numerical simulations and/or providing conceptual insight into the essential features of the dynamics. Emergent multimode quantum phenomena revealed through the lens of such reduced models can propose unique opportunities for applications while also identifying key non-classical figures of merit to serve as concrete guideposts for future experimental efforts. In this manuscript, we introduce and review our recent progress in tackling these unique challenges in broadband quantum optics.

\section{Multimode extensions of quantum input-output theory: sync-pumped parametric oscillators} \label{sec:spopo}
One of the most successful modeling frameworks for quantum systems to date has been the \emph{input-output formalism} of traditional quantum optics~\cite{Gardiner1985a, Wiseman2010}: a carefully delimited ``system''---described by a relatively simple Hamiltonian---interacts with an external ``reservoir''. For such ``open-dissipative systems'', input-output theory makes a key simplifying assumption that the reservoir, despite in many cases being a multimode and infinite-dimensional quantum field, is only ever weakly excited by the system, so its effect on the system dynamics can be assumed Markovian, allowing the use of a \emph{master equation} with simple, phenomenological Lindblad terms. While the input-output formalism has been very effective at describing discrete, nearly-closed quantum systems like single-mode anharmonic oscillators and few-atom cavity quantum electrodynamics~\cite{Mabuchi2002, Xu2015, Trived2018}, it is not immediately clear that this framework would continue to scale well when we move away from single-mode physics and into the realm of ultrafast quantum optics, where the systems are (i) highly multimode in frequency/time, and (ii) not naturally compatible with a system-bath description in the same way as cavities (consider, e.g., a ``system'' consisting of a pulse in a waveguide where the boundary conditions are technically at infinity).

Nevertheless, cavity-based ultrafast optics can be experimentally relevant, such as in the form of \emph{synchronously-pumped oscillators}, where, in addition to the usual cavity resonance condition, we impose an additional timing constraint that the cavity roundtrip time is synchronized to the repetition of a pump laser. Such oscillators are especially useful as nonlinear optical devices since the pulsed nature of the intracavity field provides an instantaneous enhancement of the peak intensity on top of the usual recycling enhancement of the cavity that results in longer effective interaction lengths~\cite{Boyd2008}. An archetype of this architecture is the synchronously-pumped optical parametric oscillator (SPOPO), which produces parametric gain and oscillation at much lower threshold power (and thus lower thermal loading) than their continuous-wave counterparts, and the broadband, ultrafast light produced by SPOPOs have a number of classical applications~\cite{Adler2009, Kippenberg2011, Spaun2016, Marandi2016}. In the quantum domain, the squeezing properties of SPOPOs~\cite{Roslund2014, Gerke2015, Shelby1992} were explored in a series of detailed studies~\cite{Valcarcel2006, Patera2010, Pinel2012}, where it was found that the interplay between the spectrum of the pump and the dispersion of the signal pulses in the SPOPO can produce rich structure in the highly-multimode entangled state of the SPOPO below threshold.

The mathematical technique used in this prior work~\cite{Valcarcel2006, Patera2010} which enabled analysis of the interactions among tens-of-thousands of frequency modes was the use of \emph{spectral supermodes}. By rewriting the quantum input-output theory of the multimode SPOPO in a supermode form, the resulting Hamiltonian of the intracavity quantum state could be truncated to a lower-dimensional representation (involving, e.g., only a few Hermite-Gaussian spectral supermodes). Building off this technique, one of our aims was to test how far a supermode-based input-output theory can take us towards a general treatment of, and simulation methods for, broadband quantum optics. In our work~\cite{Onodera2018}, we extended the below-threshold study of the SPOPO to include the construction of \emph{nonlinear Lindblad operators}, which allowed us to continue using traditional input-output theory to study the SPOPO in the \emph{above-threshold} regime. This approach fits easily into the established framework of quantum-optical master equations, where these novel nonlinear Lindblad terms are responsible for capturing the effects of nonlinear gain saturation and pump depletion which become important above threshold. We found that when the nonlinearity is large compared to the system out-coupling rate, the squeezing spectrum predicted by the below-threshold theory~\cite{Patera2010} becomes distorted due to gain saturation at and beyond threshold. Even deeper in the quantum regime of SPOPOs, these nonlinear effects, which may also be thought of as broadband two-photon loss from the perspective of the intracavity signal state, can even produce multimode Schr\"odinger cat states~\cite{Wang2016}, clearly indicating the potential for non-Gaussian physics in SPOPOs pushed into the quantum regime.

For those familiar with the physics of continuous-wave parametric oscillators, these findings perhaps come as no surprise, but we believe it is important to emphasize the intrinsic and inseparable effects that multimode physics imposes on these systems: for example, as expected, the multimode cat states populate different low-order supermodes, but one key effect of the broadband two-photon loss is to \emph{entangle} these supermodes, making the preparation of a pulsed ``single-mode'' Schr\"odinger cat state a nontrivial problem in dispersion and dissipation engineering~\cite{Brecht2015,Kienzler2015}. Such multimode phenomena are particularly important to understand because the enhanced nonlinearities afforded by ultrafast SPOPOs make them prime candidates for quantum-limited all-photonic information processing, \emph{provided} multimode effects can be sufficiently controlled. Using the phenomenology of our nonlinear SPOPO quantum model, we have recently formulated concrete design principles and experimental parameters required to reach a quasi-single-mode strong-coupling regime with nanophotonic SPOPOs on lithium niobate, and we will be reporting those results in our final version of this work to be published shortly.

\section{Continuum limit of broadband nonlinear quantum optics: Few-photon parametric downconversion and Fano physics}
\label{sec:fano-pdc}
While the example of the SPOPO discussed in Sec.~\ref{sec:spopo} both establishes the potential for ultrafast enhancement of quantum optical nonlinearities and highlights the subtleties caused by the multimode nature of broadband interactions, the input-output formalism~\cite{Xu2015, Trived2018} itself has limitations that are especially conspicuous when dealing with pulsed systems. By construction, input-output theory, even in the multimode setting, abstracts the intra-cavity dynamics into localized interactions among system modes, resulting in continuous-time dynamics per the governing Lindblad master equation. As a result, as discussed in that work~\cite{Onodera2018}, the discrete, pulsed nature of the physics is averaged out so that we only have access to system dynamics on the timescale of the cavity lifetime, which in turn needs to be larger than the roundtrip timescale. Thus, low-finesse cavities---or, indeed, waveguides with no cavity boundary conditions at all---which feature high \emph{single-pass} nonlinearity, pump-depletion, and/or loss may not be well-described by traditional input-output theory. To accurately model such systems, we need to consider the propagation and interaction of quantum fields in continuous frequency and space without reducing to a set of slowly-varying amplitudes of isolated optical (longitudinal) modes.

It turns out this field-theoretic formulation of quantum optics has already been established for the study of numerous specific phenomena such as broadband Kerr squeezing and photon bound states~\cite{Drummond1990, Drummond1987, Drummond1999_novel_solitons}. Our work in this domain of quantum optics has been strongly inspired by these continuum-limit fundamental models, whereas our central objective is to build up, from these fundamental models, an ``abstraction stack'' of numerical and modeling methodologies at varying levels of generality. As an initial foray, we discuss in this section our recent work on studying few-photon parametric downconversion in highly nonlinear waveguides~\cite{Yanagimoto2020_Fano}, where we find a striking analogy between these strongly-interacting few-photon dynamics and the physics of Fano-type discrete-continuum interactions~\cite{Fano1961, Limonov2017}. By exploiting this analogy, we not only derive complete analytic expressions for system eigenstates under this subspace of the continuum Hamiltonian, but we also discover and can intuitively explain various exotic quantum features such as complete pump depletion and Rabi-like oscillations with subpolynomial decay.

\begin{figure}[tbh]
    \centering
    \includegraphics[width=0.95\textwidth]{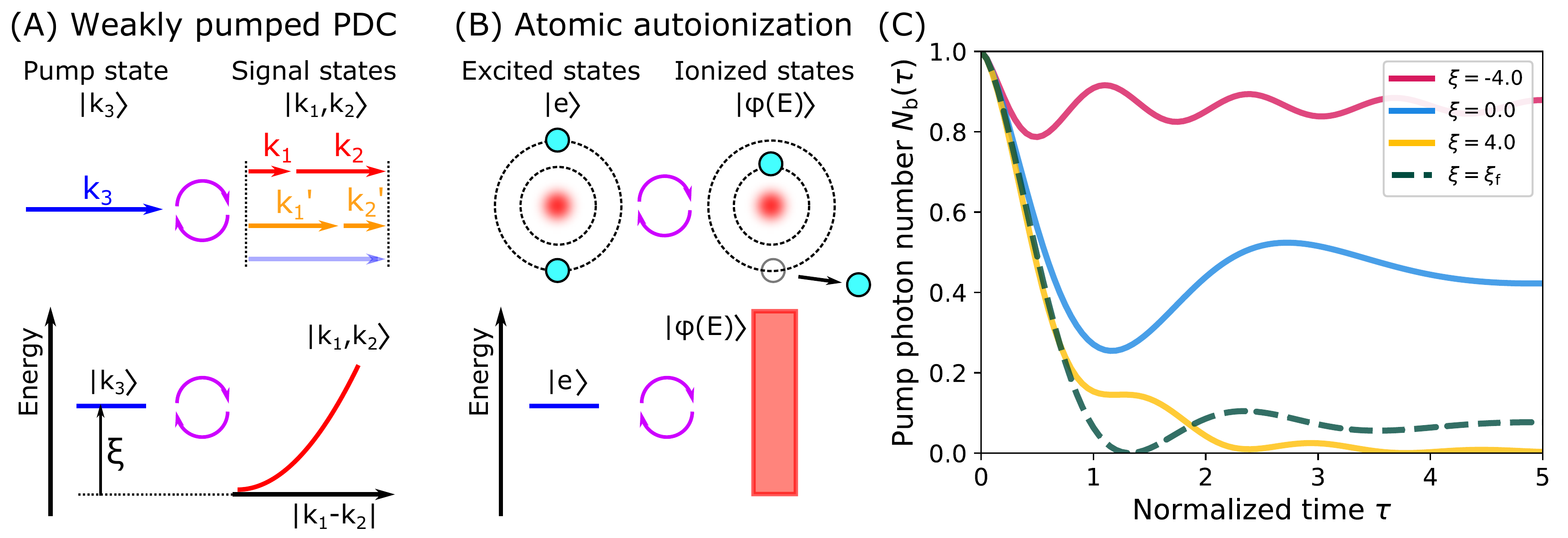}
    \caption{(A) Illustration of weakly pumped broadband PDC where a discrete pump state with momentum $\ket{k_3}$ is coupled via a $\chi^{(2)}$ nonlinear interaction to a continuum of signal states $\ket{k_1,k_2}$ with $k_1+k_2=k_3$. The energy of signal states depends quadratically on the momentum degeneracy $|k_1-k_2|$ due to quadratic energy dispersion, forming a semi-infinite energy band. The phase mismatch $\xi$ determines the energy of the pump measured from the bottom of the signal continuum. (B) In an atomic autonization, a discrete excited state $ \ket{e}$ is coupled to a continuum of ionized states $\ket{\phi(E)}$, which results in atomic ionization via a unitary process. (C) Time evolution of the pump photon population $N_\text{b}(\tau)$ for broadband single-pump-photon PDC with various phase mismatch $\xi$. Here, $\xi_\text{f}\approx \num{1.90}$ is a special value of the phase mismatch leading to complete pump depletion $N(\tau_\text{f})=0$ at finite time $\tau_\text{f}\approx\num{1.32}$.}
    \label{fig:fano}
\end{figure}

As illustrated in Fig.~\ref{fig:fano}(A), parametric downconversion (PDC)~\cite{Conteau2018,Helt2015,Christ2013,Grice1997} is a process in which a pump photon with momentum $k_3$ downconverts to a pair of signal photons with momentum $k_1$ and $k_2$ such that the total momentum is conserved $k_3=k_1+k_2$. Here, because the signal photon pair has an additional continuous degree of freedom in partitioning $k_3$, PDC can be viewed as an interaction between a discrete state $\ket{k_3}$ and a continuum of states $\ket{k_1,k_2}$. Such discrete-continuum interactions were identified by Ugo Fano in his seminal paper~\cite{Fano1961} to describe the physics of atomic autoionization, a process whereby an atom ionizes through a unitary process due to the coupling between a discrete excited state and a continuum of ionized states, as shown in Fig.~\ref{fig:fano}(B). In fact, for the case of weakly pumped PDC, the analogy becomes exact at the level of the mathematical structure of the Hamiltonian, a one-to-one correspondence which allows us to analytically write down the full state evolution of the system using Fano's theory~\cite{Fano1961}. For example, the pump photon population $N_\text{b}(\tau)$ at time $\tau$ is explicitly
\begin{align}
    N_\text{b}(\tau)=\left|\left(1+\frac{\pi}{4 \lambda_\text{M}^{3/2}}\right)^{-1}+\int^\infty_0\mathrm{d}\lambda\frac{2\sqrt{\lambda}e^{-\mathrm{i}(\lambda+\lambda_\text{M})\tau}}{4\lambda(\lambda-\xi)+\pi^2} \right|^2,
\end{align}
where $-\lambda_\text{M}$ is the energy of the ``optical meson'', a photon bound state supported by parametric interactions~\cite{Drummond1997}. Here, $\xi$ represents a normalized phase mismatch between signal and pump and corresponds to the energy of the discrete pump state measured from the bottom of the signal continuum as shown in Fig.~\ref{fig:fano}(A). Notably, $\xi$ critically determines the nature of the discrete-continuum coupling and thus the qualitative properties of PDC dynamics. When $\xi>0$, the energy of the discrete pump state lies within the energy band of the signal continuum, rendering the nature of discrete-continuum interaction ``dissipative'' (i.e., energetically on-resonance). In fact, in the limit $\xi\rightarrow\infty$, the pump photon population exhibits an monotonic exponential decay $N_\text{b}(\tau)\approx e^{-\pi\tau/\sqrt{\xi}}$, resulting in \emph{unit-efficiency PDC} or complete pump depletion. Physically this is because the downconverted signal photons disperse quickly in space suppressing backconversion, and such decay-like dynamics can be seen as a photonic analog of atomic autoionization. On the other hand, for $\xi<0$, the pump energy lies outside the energy band of the signal continuum, resulting in ``dispersive'' (i.e., energetically off-resonance) coupling. In the limit $\xi\rightarrow-\infty$, $N_\text{b}(\tau)$ exhibits a Rabi-like oscillation with sub-polynomial decay $\sim1/\sqrt{\tau}$, and due to the rapid initial decrease of oscillation amplitude, it turns out the dynamics do not converge to the conventional sinusoidal oscillation of single-mode PDC~\cite{Drobny1994, Bandilla2000} in any limit, highlighting the intrinsically multimode nature of broadband parametric interactions. In Fig.~\ref{fig:fano}(C), we show analytic trajectories of $N_\text{b}(\tau)$ which highlight the critical role the phase-mismatch $\xi$ plays in determining quantum PDC dynamics. For finite $\xi$, the pump photon number takes a finite positive value even in the limit of infinite interaction time, i.e., $\lim_{\tau\rightarrow\infty}N_\text{b}(\tau)>0$; such quantum limitations to the PDC efficiency are attributed to the presence of photon bound states in a broadband medium.

These findings echo the well-known fact that phase-mismatch is a critical parameter in nonlinear optics, but with the somewhat surprising conclusion that unit-efficiency PDC occurs in the limit of \textit{large} phase mismatch, contrary to conventional physical intuition that such interactions are most efficient at resonance. Furthermore, the Fano analogy is surprisingly general and can provide intuition about physics on a ``system'' level as well~\cite{Yanagimoto2020_Fano}: for a pair of coupled nonlinear waveguides, Fano interferences among multiple discrete-continuum interactions can lead to characteristic Fano lineshapes and the formation of a bound state in the continuum~\cite{Hsu2016}. Finally, although the phenomenology of broadband few-photon PDC is not intrinsically a \emph{pulsed} effect---that is to say, the Fano-type physics are not fundamentally different for pulsed vs.\ continuous-wave inputs~\cite{Yanagimoto2020_Fano}---such exotic dynamics only arise by virtue of the highly multimode/broadband nature of the signal field in this continuum limit.

\section{Full quantum simulation with matrix product states: general pulse propagation in 1D nonlinear waveguides}
\label{sec:mps}
The model reductions employed for the work reviewed in Sec.~\ref{sec:spopo} and Sec.~\ref{sec:fano-pdc} provided useful physical insight into a variety of basic broadband nonlinear phenomena. On the numerical simulation end, however, it is useful to also develop general out-of-the-box techniques that can be deployed to perform full quantum simulations for entire classes of nonlinear broadband devices. Of course, full quantum simulation of highly multimode, fully entangled states is in general prohibitively (i.e., exponentially) expensive~\cite{Zhong2020, Arute2019, Nielsen2000, Boixo2018}, but the complexity of quantum simulations generically arises from three different sources: (i) having a large number of modes, (ii) quantum non-Gaussianity, and (iii) strong long-range entanglement. Without (i) a full numerical diagonalization of the low-dimensional system Hamiltonian is feasible. Without (ii)~\cite{Lloyd1999}, second-order correlations suffice to completely characterize the quantum state~\cite{Olivares2012, Braunstein2005}, and as a result, the system is also computationally tractable. While both of these cases have been extensively explored in quantum optics research~\cite{Asavarant2019, Wang2016, Birnbaum2005}, we immediately see that quantum dynamics in a generic broadband nonlinear waveguide simultaneously violates both of these two conditions. Crucially, however, there is another corner to this cube: the fact that the interactions in a 1D waveguide are \emph{local} implies that condition (iii) does not necessarily hold.

\begin{figure}[bth]
    \centering
    \includegraphics[width=0.98\textwidth]{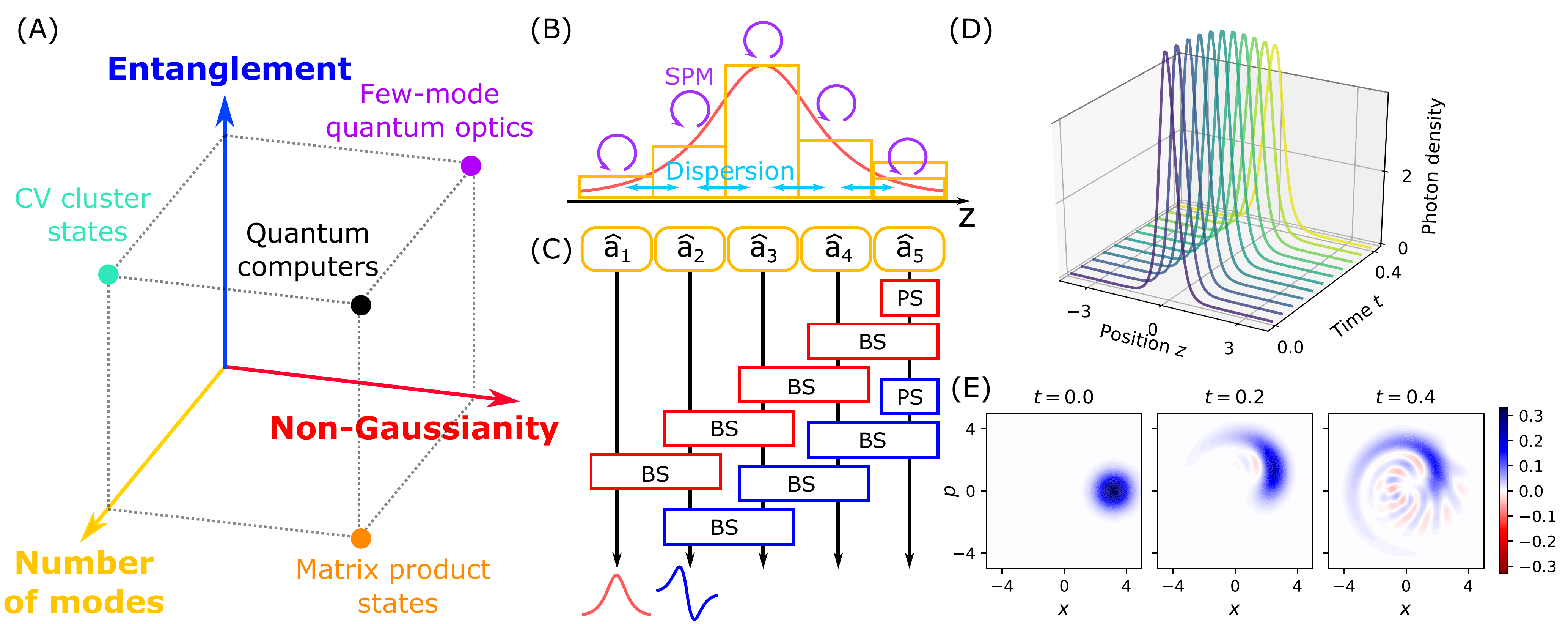}
    \caption{(A) An illustration of the different facets which contribute to the complexity of simulating continuous-variable quantum systems. Representative states/systems~\cite{Asavarant2019, Wang2016, Afek2010, Schollwoeck2011} belonging to each corner of the cube (marked by circles) are shown. (B) Quantum pulse propagation through a nonlinear medium is modeled as interacting quantum fields on discretized spatial bins. (C) Cascaded applications of beamsplitter (BS) and phase-shifter (PS) operations on an MPS can demultiplex nonlocal supermodes into local spatial bins. (D) Time evolution of the photon density distribution of a coherent-state Kerr soliton in a $\chi^{(3)}$ nonlinear waveguide. (E) Dynamics of the Wigner function of the solitonic mode. Simulations for (D) and (E) were performed using time-evolving block decimation based on MPS~\cite{Vidal2003}.}
    \label{fig:mps}
\end{figure}

In our recent work~\cite{Yanagimoto2021_mps}, we presented a numerical method that exploits this lack of complex entanglement in 1D waveguides by adapting a powerful concept in many-body physics, the matrix product state (MPS) formalism~\cite{Schollwoeck2011}. It is heuristically known that the amount of entanglement in a 1D quantum system is generically limited and local~\cite{Vidal2004}, and in such cases, the MPS representation efficiently describes the quantum state as a product of low-rank tensors. In the context of many-body physics, the use of MPS has led to the huge success of density matrix renormalization group (DMRG) in studying low-energy states of 1D systems, but Vidal in his seminal paper~\cite{Vidal2003} showed that it is also possible to directly evolve an MPS in time under the system Hamiltonian by decomposing the Hamiltonian into two-body interactions, each of which can be easily applied to an MPS. Recently, these MPS-based approaches have made their way into the realm of optics as well, where they have helped uncover rich quantum dynamics in optical systems~\cite{Lubasch2018, Manzoni2017, Mahmoodian2020}. 

However, the origin of the MPS framework in many-body physics means that such formalisms focus on the ``particle'' aspects of the field~\cite{Muth2010, Muth2010_b} as encoded by quantities such as $n$-particle correlation functions. On the other hand, in the context of pulsed wave optics, it is often more natural to think of optical fields as ``oscillator modes'' and represent the system dynamics in a temporal/frequency supermode basis~\cite{Brecht2015, Fabre2020}, as is commonly done for continuous-variable quantum systems~\cite{Braunstein2005}. To make information about nonlocal pulse supermodes accessible, we develop a ``demultiplexing'' algorithm to manipulate nonlocal information in an MPS~\cite{Yanagimoto2021_mps}. In our scheme, as shown in Fig.~\ref{fig:mps}(C), cascaded applications of one-mode and two-mode operations re-encode information about supermodes, originally distributed across the entire system, into localized bins from which that information can be efficiently extracted. Combining this demultiplexing algorithm with MPS-based dynamical simulation, we can construct a full numerical picture for the non-classical phase-space dynamics~\cite{Cahill1969, Drummond1980_p}. These techniques can, in principle, be applied to any 1D nonlinear photonic system and can also readily account for effects of decoherence induced by optical loss via, e.g., quantum trajectory theory~\cite{Wiseman2009}. Thus, the approach enables not only characterization of realistic experimental devices but also an exploration of rich physics arising fundamentally from dissipation~\cite{Bender2007, El-Ganainy2018}.

Using this framework, we studied the quantum propagation of a Kerr soliton in a 1D $\chi^{(3)}$ waveguide, which has been extensively studied in the classical regime~\cite{Agrawal2019, Kivshar2003}, while certain quantum-induced features such as waveform dispersion~\cite{Lai1989_b}, soliton evaporation~\cite{Villari2018}, and squeezing~\cite{Carter1987, Haus1990} are known to arise in the highly nonlinear regime. In Fig.~\ref{fig:mps}(D), we show how the pulse waveform of the Kerr soliton evolves in time. In the few-photon quantum regime, we find that, contrary to its classical stability, the envelope experiences a ``quantum-induced'' dispersion caused by the fact that the classical (coherent-state) soliton is not an eigenstate of the system Hamiltonian. Furthermore, established approximations for the physics of quantum Kerr solitons, such as time-dependent Hartree-Fock~\cite{Haus1989}, suggest it may be possible to view the solitonic supermode as experiencing single-mode Kerr dynamics in phase space, which would allow an initial coherent-state soliton to evolve into a non-classical state~\cite{Wright1991, Yanagimoto2020, Korolkova2001}. As shown in Fig.~\ref{fig:mps}(E), our exact quantum simulations of this system using MPS indeed produces a highly non-classical state with Wigner function negativities~\cite{Kenfack2004}, a manifestation of \textit{coherent} non-Gaussian dynamics. However, we find structural derivations from pure Kerr dynamics in the phase-space portrait, as well as loss of purity due to the entanglement of the soliton with higher-order supermodes. This refinement in our understanding of quantum Kerr soliton dynamics could prove useful in harnessing these ultrafast coherent dynamics as high-bandwidth, all-optical resources for quantum information and engineering~\cite{Yanagimoto2020}.

\section{Conclusions and outlook}
\label{sec:conclusion}
Quantum models of broadband nonlinear-optical phenomena are inherently high-dimensional.  With the advent of new experimental capabilities for fabricating high-quality nanophotonic structures in materials such as lithium niobate, and for performing quasi-phase matching and dispersion engineering to enable the use of ultrafast pump and signal fields, we need to develop new theoretical approaches to deriving accurate reduced models that capture the essential physics of such systems while facilitating intuitive conceptual analysis and device/system design.  There are indications that some core methods of traditional quantum optics, such as conventional input-output theory, may need to be generalized substantially to apply in the strong coupling regime of ultrafast nanophotonics. Early explorations of quantum model reduction for ultrafast nanophotonics highlight a key role for dispersion engineering and related phase-matching considerations in isolating small subsets of supermodes on which coherent quantum dynamics occur. By pursuing a deeper understanding of both the generality and limitations of such reduced models, we ultimately hope to identify core physical concepts and modeling methodologies to comprise a ``system-process-control'' paradigm for quantum engineering on this burgeoning hardware platform.

\section*{Funding}
Army Research Office (W911NF-16-1-0086); National Science Foundation (CCF-1918549, PHY-2011363).
\section*{Acknowledgments}
The authors wish to thank NTT Research for their financial and technical support. R.\,Y.\ is supported by a Stanford Q-FARM Ph.D.\ Fellowship and the Masason Foundation. 

\bibliography{report} 
\bibliographystyle{spiebib} 

\end{document}